# Constraints on Tsallis Cosmology from Big Bang Nucleosynthesis and the Relic Abundance of Cold Dark Matter Particles

Petr Jizba [1]  and Gaetano Lambiase [2,3,*]

1 Faculty of Nuclear Sciences and Physical Engineering, Czech Technical University in Prague, Břehová 7, 115 19 Praha, Czech Republic; p.jizba@fjfi.cvut.cz

2 Dipartimento di Fisica "E.R. Caianiello", Universita' di Salerno, I-84084 Fisciano, SA, Italy

3 INFN—Gruppo Collegato di Salerno, I-84084 Fisciano, SA, Italy

* Correspondence: lambiase@sa.infn.it

**Abstract:** By employing Tsallis' extensive but non-additive $\delta$-entropy, we formulate the first two laws of thermodynamics for gravitating systems. By invoking Carathéodory's principle, we pay particular attention to the integrating factor for the heat one-form. We show that the latter factorizes into the product of thermal and entropic parts, where the entropic part cannot be reduced to a constant, as is the case in conventional thermodynamics, due to the non-additive nature of $S_\delta$. The ensuing two laws of thermodynamics imply a Tsallis cosmology, which is then applied to a radiation-dominated universe to address the Big Bang nucleosynthesis and the relic abundance of cold dark matter particles. It is demonstrated that the Tsallis cosmology with the scaling exponent $\delta \sim 1.499$ (or equivalently, the anomalous dimension $\Delta \sim 0.0013$) consistently describes both the abundance of cold dark matter particles and the formation of primordial light elements, such as deuterium $^2H$ and helium $^4He$. Salient issues, including the zeroth law of thermodynamics for the $\delta$-entropy and the lithium $^7Li$ problem, are also briefly discussed.

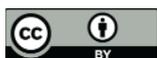

## 1. Introduction

Since the discovery of black hole thermodynamics by Bekenstein [1] and Hawking [2], it has become clear that a non-perturbative analysis of Einstein's gravity could potentially be linked to holographic thermodynamics. In particular, Jacobson's pioneering work [3] has been instrumental in pointing out a deep formal connection between holographic thermodynamics and gravitation, culminating in the derivation of the Einstein field equations (see also Refs. [4–6]) and the cosmological equations (the Friedmann equations) [7–18] from the first two laws of thermodynamics. These studies have generated considerable interest, paving the way for an analysis that would extend to a broader range of entropies than just the standard Boltzmann–Gibbs–Shannon entropy [19–29]. Ensuing models generally account for various modifications of the Bekenstein–Hawking (BH) entropy area law where the conventional holographic scenario is inapplicable. For example, when considering entropic cosmology [30] or loop quantum gravity [31–33], logarithmic corrections to the area law are observed as a result of entanglement between quantum fields situated both within and beyond the horizon [34–38]. Similarly, generalized non-additive entropies typically tend to induce a power-law behavior rather than an area law; see, e.g., Refs. [19,39–42].

It should be stressed that the opposite behavior is also true; namely, the area-law formula for black hole entropy holds only in Einstein's theory, i.e., when the ensuing action functional includes only a linear term of the scalar curvature $R$. For instance, the Bekenstein–Hawking entropy no longer holds in generic higher-derivative gravity theories [12], and, in particular, in $f(R)$ gravity, the entropy of a static black hole acquires the form $S \propto L^2 f'(R)$, cf., e.g., Ref. [43].

Recently, C. Tsallis proposed a thermodynamic entropy in 3 spatial dimensions for systems with sub-extensive scaling of microstates, such as, e.g., black holes. This so-called $\delta$-entropy is an entropic functional of the form [19,42]

$$S_\delta = \eta_\delta \sum_i p_i \left(\log \frac{1}{p_i}\right)^\delta, \quad \delta > 0. \tag{1}$$

Here, the values $p_i$ represent the probabilities of elementary events (or microstates), and the multiplicative constant $\eta_\delta$ reflects the units used to measure the entropy. An equiprobable distribution (1) acquires the form

$$S_\delta = \eta_\delta (\log W)^\delta, \tag{2}$$

where $W$ is a number of available microstates. According to Ref. [19], the entropy (1) can be regarded as a valid thermodynamic entropy in 3 spatial dimensions for systems with the sub-extensive scaling. Such a scaling typically appears in various holographic scenarios. For instance, according to the holographic principle, the entropy of a black hole and more generally the entropy of the universe is a Shannon entropy with a peculiar area-law scaling, namely

$$S_{\text{BH}} = -k_B \sum_i p_i \log p_i \propto L^2. \tag{3}$$

Here, $L$ is a characteristic length-scale in the problem, and the Boltzmann constant $k_B$ is typically chosen in the context of holographic thermodynamics. By the asymptotic equipartition property [44], the $S_{\text{HB}}$ entropy behaves as

$$S_{\text{BH}} \propto \log W, \tag{4}$$

implying that the number of microstates $W$ (more precisely a volume of a typical set) scales exponentially so that

$$W = f(L)\eta^{L^2}, \quad \text{with} \quad \eta > 1 \quad \text{and} \quad \lim_{L \to \infty} f(L)/L = 0. \tag{5}$$

While the scaling (5) prevents the Bekenstein–Hawking entropy from being considered as a full-fledged thermodynamic entropy, the entropy (1) may be considered a proper thermodynamic entropy, provided a suitable scaling exponent $\delta$ is chosen [19]. This is because with a proper $\delta$, the entropy $S_\delta$ preserves the structure of thermodynamic Legendre transforms [19,41,42].

By combining (2) with (5), we obtain that in the large $L$ limit, the entropy $S_\delta$ can be written as

$$S_\delta = \gamma_\delta A^\delta, \tag{6}$$

where $A$ is the horizon area and $\gamma_\delta$ is a $\delta$-dependent constant, which for $\delta = 1$ reduces to Hawking's conventional form $\gamma = 1/(4L_p^2)$. When the number of microstates scales according to (5), the scaling exponent $\delta$ should be 3/2 in three spatial dimensions to ensure the entropy is an extensive thermodynamic quantity [19,42]. On the other hand, it is expected that at the quantum gravitational level, the black hole surface (and by extension cosmological horizons) will follow a deformed holographic scaling [45–47].

In particular, one may employ Barrow's idea, which posits that black holes and more generally cosmological horizon surfaces possess a fractal structure with associated generalized Bekenstein–Hawking entropy

$$S_{\text{Gen.BH}} \propto L^{2+\Delta}. \tag{7}$$

Here, $\Delta$ is nothing but an *anomalous dimension* because similarly as in conventional Quantum Field Theory (QFT), it simply measures how much the scaling dimension (i.e., $2 + \Delta$) deviates from its classical value (i.e., 2) due to quantum effects. Since the coupling constant in conventional quantum gravity decreases with increasing distance, the larger the distance scale [48], the smaller the value of $\Delta$. At large scales (low energies), it may be expected that $\Delta = 0$, and one recovers the classical Bekenstein–Hawking entropy. Equation (7) implies the scaling

$$W = f(L)\eta^{L^{2+\Delta}}, \quad \text{with} \quad \eta > 1. \tag{8}$$

The Barrow entropy (7) was originally proposed in Ref. [45] as a toy model for understanding the possible effects of quantum gravitational spacetime foam [49]. The Barrow entropy reduces to the standard Bekenstein–Hawking entropy in the limit $\Delta \to 0$, whereas the case $\Delta = 1$ corresponds to maximal deformation. Barrow provided a simple "sphere-flake" fractal model for $\Delta$, which allows only for $\Delta \in [0, 1]$. While indeed the Hausdorff dimension of very rough surfaces might be arbitrarily close to the embedding Euclidean dimension (i.e., max 3), the lower value of $\Delta$ might acquire negative values for "spongy" or "porous" surfaces. For instance, for the Sierpinski carpet, the Hausdorff dimension is ~1.89, so that $\Delta \sim -0.11$, while real-world porous surfaces can have Hausforff dimensions substantially lower than 2, cf., e.g., [50,51]. The fact that anomalous dimensions can be negative stems also from QFT where the renormalization group reasonings generally allow for negatively-valued $\Delta$ in various systems [52,53]. In passing, we might note that for $\Delta > 0$, the scaling (8) indicates that there are more quantum microstates available than there are in the classical situation, while for $\Delta < 0$, the number of available states is lower than what is seen classically.

If we now insert Barrow's microstate scaling (8) to (2), we obtain

$$S_\delta = \gamma_\delta A^{(1+\Delta/2)\delta}. \tag{9}$$

The extensivity of the $\delta$-entropy then implies the relation between $\Delta$ and $\delta$, namely

$$(1 + \Delta/2)\delta = \frac{3}{2} \quad \Leftrightarrow \quad \delta = \frac{3}{2+\Delta}. \tag{10}$$

Entropy $S_\delta$ belongs to the two-parameter class of entropic functionals, referred to as the $S_{q,\delta}$ entropies that were proposed by Tsallis in Ref. [19]. There, in particular, $S_\delta \equiv S_{1,\delta}$. It is important to stress that $S_\delta$ does not correspond to the widely used Tsallis entropy [54–57], which is a prominent concept in statistical physics and the theory of complex dynamical systems. Specifically, Tsallis' entropy with the non-extensivity parameter $q$ is the $S_{q,1}$ member in the aforementioned two-parameter class of entropies. The so-called *Tsallis cosmology* is an approach that incorporates the $S_\delta$ entropy into the first law of thermodynamics to produce modified Friedmann cosmological equations. The standard cosmological model is then recovered in the limit $\delta = 1$ and $\Delta = 0$. It is worth noting that the use of $S_\delta$ in formulating the first law of thermodynamics appears to be somewhat arbitrary in the existing literature. For this reason, we adhere to Tsallis' original suggestion [19] for $S_\delta$ in this paper and formulate the first law so that the entropy will be extensive but not additive. Due to the non-additive nature of the entropy, particular attention must be paid to the integration factor of the heat one-form, which in this case is not a simple inverse of thermodynamic temperature, but instead, it factorizes into entropic and thermal parts.

With the correct first law of thermodynamics at hand, one can explore the potential consequences of Tsallis cosmology. It is clear that the consistency of the approach strictly relies on the available observational datasets that should be matched with Tsallis cosmology at various epochs. In particular, the Big Bang nucleosynthesis (BBN) plays a crucial role in this respect, providing an independent and powerful constraint for any cosmological model. Notably, the formation of primordial light elements in the BBN represents an important epoch of the universe's evolution. In fact, during this era, the formation of light elements left an imprint on their abundance that is observed today. With the advancements in high-precision instrumentation and the infusion of new ideas from particle cosmology and astrophysics, the BBN currently represents a powerful probe for testing the early universe, with non-trivial consequences on scenarios beyond the Standard Models of particle physics and the standard cosmological model. Ensuing "new" physics may alter the evolution of events that occurred in the BBN era as compared to the standard theories, and current observations thus provide strong constraints on parameters characterizing such models. One of the key messages of this paper is that Tsallis cosmology is capable of consistently describing the primordial light elements' formation in the BBN era. In addition, we also show that the range of $\delta$-parameters obtained is compatible with bounds on the cold dark matter relic abundance.

In contrast to other works on cosmology that rely on the $\delta$-entropy, we put emphasis on the second law of thermodynamics. By the second law of thermodynamics, we mean the Carathéodory formulation. This formulation states that the heat one-form in any thermodynamically consistent system should be holonomic, which implies the existence of a new state function—entropy. Moreover, this allows for the definition of a unique absolute temperature for the $\delta$-entropy-driven thermodynamic systems. We demonstrate that the integration factor of the heat one-form cannot simply be described as the inverse of thermodynamic temperature due to the non-additive nature of entropy. Instead, it factorises into entropic and thermal parts. We further show that the factorization property of the integration factor allows us to identify absolute temperature uniquely (up to a multiplicative factor). This, in turn, permits us to follow the established methodology from conventional thermodynamics to derive the (modified) Friedmann equations. It is worth noting that our use of $S_\delta$ is based on Formula (9), which conceptually differs from a more commonly used version [58–60]. Therefore, while the first law of thermodynamics reflects energy conservation, and as such, it is crucial in setting up the Friedmann equations (basically along the same lines as in the original Jacobson's paper [3]), it is the second law (more precisely its modified version—new integration factor, new entropy, and new absolute temperature) that brings about the key modifications into the Friedmann equations and allows for novel cosmological implications.

The layout of the paper is as follows. In the following section, we examine the role of the $S_\delta$ entropy and discuss how it can be incorporated into the thermodynamic framework. Particular attention is paid to an integrating factor for the heat one-form. It is shown that the latter cannot be simply identified with the inverse thermodynamic temperature, but instead, it factorises into entropic and thermal parts. In Section 2.2, we briefly discuss the modified Friedmann equations that result from the application of the first law of thermodynamics to the apparent horizon of a FRW (Friedmann–Robertson–Walker) universe. We use the constraints from the BBN physics in Section 3 and from dark matter in Section 4 to infer limits on the anomalous dimension and on Tsallis' parameter $\delta$. It is shown that the Tsallis cosmology is consistent with both the formation of primordial light elements (such as deuterium $^2H$ and helium $^4He$) and the relic abundance of dark matter particles given that the scaling exponent $\delta \sim 1.499$, or equivalently, the anomalous dimension $\Delta \sim 0.013$. Finally, Section 5 summarizes our results and identifies potential avenues for future research. For the sake of clarity, we relegate some more technical considerations to two appendices.

## 2. Thermodynamics Based on $S_\delta$ and Cosmological Equations

### 2.1. $S_\delta$-Entropy and the First and Second Laws of Thermodynamics

In this section, we briefly review thermodynamics based on the $S_\delta$-entropy. In particular, we will focus on introducing the $S_\delta$-entropy into the first law of thermodynamics by utilizing Carathéodory's formulation of the second law of thermodynamics. It is important to note that while Carathéodory's formulation may not be as common in the literature, it can be derived directly from the conventional Kelvin–Planck statement of the second law [61,62]. In our exposition, we will loosely follow Ref. [41]. The result obtained will be further instrumental in deducing modified Friedmann–Robertson–Walker equations, which will be discussed in following sub-section.

Many cosmological systems, such as black holes, have entropies that exhibit sub-extensive scaling. A paradigmatic example of this phenomenon is the area-law scaling of the BH entropy. Since the laws of black hole mechanics are mathematically analogous to the laws of thermodynamics, one often formally postulates black hole thermodynamics without any reference to arguments coming from statistical mechanics [63]. This strategy was further extended by Gibbons and Hawking [64] and later by 't Hooft and Susskind [65,66], who have demonstrated that black hole thermodynamics is more general than black holes, namely that cosmological event horizons also have an entropy and temperature and that one may again affiliate formal thermodynamic rules with them. These findings have prompted an ongoing debate on whether the aforementioned systems are merely analogous to thermodynamic systems or whether they should be considered genuine thermodynamic systems. Recently, Tsallis offered an alternative viewpoint [19,42] in which he advocated that such systems may be viewed as genuine thermodynamic systems, provided the cosmological entropy is replaced with an extensive but not additive entropy, while the holographic scaling of the state-space remains unchanged.

Let us now take a closer look at Tsallis' proposal. We start by recalling that the key property in the thermodynamic framework is the Legendre transform, which, for instance, for the Gibbs free energy, takes the form

$$G(T, p, N, \ldots) = U(S, V, N, \ldots) + pV - TS, \tag{11}$$

where $G$ and $U$ stand for the Gibbs free energy and internal energy, respectively. Both $G$ and $U$ are expressed in terms of their *natural variables*, and dots stand for prospective additional (non-mechanical) state variables.

By following [19,42], we now define the length-scale independent thermodynamic potentials $g = \lim_{L \to \infty} G/L^\varepsilon$ and $u = \lim_{L \to \infty} U/L^\varepsilon$, where $L$ is the characteristic linear scale of the system and $\varepsilon$ is a scaling exponent (not necessarily identical to the spatial dimension $d$). Note that $g$ and $u$ must satisfy (for large $L$)

$$G(T, p, N, \ldots) = L^\varepsilon g(T/L^\nu, p/L^\nu, N/L^d, \ldots),$$
$$U(S, V, N, \ldots) = L^\varepsilon u(S/L^d, 1, N/L^d, \ldots). \tag{12}$$

Here, we do not assume that the scaling exponent $\nu$ has a typical laboratory value $\nu = 0$. Therefore, because (for large $L$) $G(T, p, N, \ldots) \propto L^\varepsilon$, $U(S, V, N, \ldots) \propto L^\varepsilon$, $p \propto L^\nu$, and $T \propto L^\nu$, then (11) inevitably implies that $S \propto L^d$ (this was implicitly used in (12)) and $\varepsilon = \nu + d$. In this way, one can (for large $L$) rewrite (11) in the form

$$g(T/L^\nu, p/L^\nu, N/L^d, \ldots) = u(S/L^d, 1, N/L^d, \ldots) + \frac{p}{L^\nu} \cdot 1 - \frac{T}{L^\nu} \frac{S}{L^d}, \tag{13}$$

Hence, the structure of the Legendre transform is also satisfied for length-scale-independent thermodynamic potentials. This analysis shows that entropy should be an extensive quantity provided that $T$ and $p$ scale in the same manner, regardless of the precise scaling of the thermodynamic potentials (which should be the same for all of them as they all refer to energy). In addition, it is clear that one could also repeat the same reasoning for other thermodynamic potentials. The required extensivity of thermodynamic entropy is a starting point of Tsallis' analysis. In order to satisfy both the holographic state-space scaling (5) (and more generally (7)) and the extensivity condition $S \propto L^d$, Tsallis proposed the $S_\delta$ entropy with a specific value of $\delta$ that enforces the extensivity. In particular, for $d = 3$, one has that $\delta$ should be $3/2$ for conventional holographic scaling (5) and $3/(2 + \Delta)$ for Barrow's type of scaling (8).

In the spirit of Tsallis' suggestion, we now set $\alpha = 2 + \Delta$ and assume that $S_{3/\alpha}$ is a thermodynamic entropy. There are two apparent drawbacks associated with this assumption. First, $S_{3/\alpha}$ is not additive (not even in the $L \to \infty$ limit) but instead follows the pseudo-additivity rule

$$S_{3/\alpha}(A + B) = \left[ S_{3/\alpha}^{\alpha/3}(A) + S_{3/\alpha}^{\alpha/3}(B) \right]^{3/\alpha}, \tag{14}$$

for any two independent subsystems $A$ and $B$. Second, it is unclear what *conjugate thermodynamic variable* is associated with this entropy.

The first point, which is an unavoidable result of working with systems that have sub-extensive scaling, such as gravity, is not a major issue, as we shall see. However, the second point is more serious. Carathéodory's formulation of the second law of thermodynamics states that a heat one-form, $đ\mathcal{Q}$, must have an integration factor (with the heat one-form being *holonomic*) so that entropy is a state function [67,68]. However, since the entropy is not additive, one cannot use the conventional Carnot cycle argument [69] in the proof of Clausius equality to simply equate the integration factor with inverse temperature. Let us examine this last point more closely to understand better what is involved. Since the exact differential associated with the heat one-form is entropy, we can write

$$dS_{3/\alpha}(\mathbf{a}, \vartheta) = \mu(\mathbf{a}, \vartheta) \, đ\mathcal{Q}(\mathbf{a}, \vartheta), \tag{15}$$

where $\mathbf{a}$ represents a collection of relevant state variables and $\vartheta$ is some *empirical* temperature whose existence is guaranteed by the zeroth law of thermodynamics (see also Appendix A). We now divide the system in question into two subsystems, $A$ and $B$, that are described by state variables $\{\mathbf{a}_1, \vartheta\}$ and $\{\mathbf{a}_2, \vartheta\}$, respectively. Then,

$$đ\mathcal{Q}_A(\mathbf{a}_1, \vartheta) = \frac{1}{\mu_A(\mathbf{a}_1, \vartheta)} dS_{A,3/\alpha}(\mathbf{a}_1, \vartheta),$$

$$đ\mathcal{Q}_B(\mathbf{a}_2, \vartheta) = \frac{1}{\mu_B(\mathbf{a}_2, \vartheta)} dS_{B,3/\alpha}(\mathbf{a}_2, \vartheta). \tag{16}$$

Therefore, for the whole system

$$đ\mathcal{Q}_{A+B} = đ\mathcal{Q}_A + đ\mathcal{Q}_B, \tag{17}$$

with

$$đ\mathcal{Q}_{A+B}(\mathbf{a}_1, \mathbf{a}_2, \vartheta) = \frac{1}{\mu_{A+B}(\mathbf{a}_1, \mathbf{a}_2, \vartheta)} dS_{(A+B),3/\alpha}(\mathbf{a}_1, \mathbf{a}_2, \vartheta), \tag{18}$$

we can write

$$dS_{(A+B),3/\alpha}(\mathbf{a}_1, \mathbf{a}_2, \vartheta) = \frac{\mu_{A+B}(\mathbf{a}_1, \mathbf{a}_2, \vartheta)}{\mu_A(\mathbf{a}_1, \vartheta)} dS_{A,3/\alpha}(\mathbf{a}_1, \vartheta)$$

$$+ \frac{\mu_{A+B}(\mathbf{a}_1, \mathbf{a}_2, \vartheta)}{\mu_B(\mathbf{a}_2, \vartheta)} dS_{B,3/\alpha}(\mathbf{a}_2, \vartheta). \tag{19}$$

Let us now assume that there is only one state variable so that $\mathbf{a} = a$. If there were more state variables, our subsequent argument would still be valid, but we would need to consider more than two subsystems. Under this assumption, we can invert $S_{A,3/\alpha}(a_1, \vartheta)$ and $S_{B,3/\alpha}(a_b, \vartheta)$ in terms of $a_1$ and $a_2$ and write (at least locally) that

$$a_1 = a_1(S_{A,3/\alpha}, \vartheta) \quad \text{and} \quad a_2 = a_2(S_{B,3/\alpha}, \vartheta). \tag{20}$$

With this, Equation (19) can be cast into the form

$$
\begin{aligned}
dS_{(A+B),3/\alpha}(S_{A,3/\alpha}, S_{B,3/\alpha}, \vartheta) &= \frac{\mu_{A+B}(S_{A,3/\alpha}, S_{B,3/\alpha}, \vartheta)}{\mu_A(S_{A,3/\alpha}, \vartheta)} dS_{A,3/\alpha} \\
&+ \frac{\mu_{A+B}(S_{A,3/\alpha}, S_{B,3/\alpha}, \vartheta)}{\mu_B(S_{B,3/\alpha}, \vartheta)} dS_{B,3/\alpha} + 0 d\vartheta.
\end{aligned}
\tag{21}
$$

Since $dS_{3/\alpha}$ is a total differential, the following integrability conditions must hold:

$$
\frac{\partial \log(\mu_A(S_{A,3/\alpha}, \vartheta))}{\partial \vartheta} = \frac{\partial \log(\mu_B(S_{B,3/\alpha}, \vartheta))}{\partial \vartheta} = \frac{\partial \log(\mu_{A+B}(S_{A,3/\alpha}, S_{B,3/\alpha}, \vartheta))}{\partial \vartheta},
\tag{22}
$$

$$
\frac{1}{\mu_A(S_{A,3/\alpha}, \vartheta)} \frac{\partial \mu_{A+B}(S_{A,3/\alpha}, S_{B,3/\alpha}, \vartheta)}{\partial S_{B,3/\alpha}} = \frac{1}{\mu_B(S_{B,3/\alpha}, \vartheta)} \frac{\partial \mu_{A+B}(S_{A,3/\alpha}, S_{B,3/\alpha}, \vartheta)}{\partial S_{A,3/\alpha}}.
\tag{23}
$$

Note that in (22), the derivatives cannot depend on entropy but only on $\vartheta$. We can thus denote the right-hand-side (RHS) of (22) as $-w(\vartheta)$ and write the solutions in the form

$$
\mu_A(S_{A,3/\alpha}, \vartheta) = \Phi_A(S_{A,3/\alpha}) \exp\left(-\int w(\vartheta) d\vartheta\right) = \Phi_A(S_{A,3/\alpha}) T^{-1}(\vartheta),
$$

$$
\mu_B(S_{B,3/\alpha}, \vartheta) = \Phi_B(S_{B,3/\alpha}) \exp\left(-\int w(\vartheta) d\vartheta\right) = \Phi_B(S_{B,3/\alpha}) T^{-1}(\vartheta),
$$

$$
\mu_{A+B}(S_{A,3/\alpha}, S_{B,3/\alpha}, \vartheta) = \Phi_{A+B}(S_{A,3/\alpha}, S_{B,3/\alpha}) \exp\left(-\int w(\vartheta) d\vartheta\right)
$$

$$
= \Phi_{A+B}(S_{A,3/\alpha}, S_{B,3/\alpha}) T^{-1}(\vartheta).
\tag{24}
$$

Here, $\Phi_X$ (where $X$ stands for $A$, $B$, and $A + B$, respectively) are some arbitrary functions of the entropy, and $T(\vartheta)$ is a subsystem-independent (but generally $\alpha$-dependent) function of the empirical temperature. The negative sign in front of $w(\vartheta)$ is adopted to ensure that for the monotonically increasing function $w(\vartheta)$, the temperature function $T(\vartheta)$ will be a monotonically increasing function of the empirical temperature, $\vartheta$.

By differentiating Equation (14), we can observe that

$$
dS_{(A+B),3/\alpha} = \frac{S_{A,3/\alpha}^{\alpha/3-1}}{S_{(A+B),3/\alpha}^{\alpha/3-1}} dS_{A,3/\alpha} + \frac{S_{B,3/\alpha}^{\alpha/3-1}}{S_{(A+B),3/\alpha}^{\alpha/3-1}} dS_{B,3/\alpha}.
\tag{25}
$$

By comparing this with (21) and (24), we can infer that the condition $\Phi_X(S_{X,3/\alpha}) = \kappa S_{X,3/\alpha}^{1-\alpha/3}$ must hold (here, $\kappa$ is an arbitrary multiplicative constant). Using this identification, we can easily verify that the remaining integrability condition (23) is also satisfied. In conventional thermodynamics, $\Phi_X$ is a constant, enabling the integration factor to be identified with an absolute temperature. In the context of non-additive entropy $S_{3/\alpha}$, this is not the case. Fortunately, $\mu_X$ satisfies a simple factorization rule in which the dependence of $\mu_X$ on $S_{3/\alpha}$ and $\vartheta$ is separated. We note that up to a multiplicative constant (that sets the units), $T$ is a unique temperature quantifier of the system that is described by $\vartheta$. For this reason, we can identify $T$ with an *absolute temperature* (see also Appendix A). Finally, the heat one-form, $dQ$, which is part of the first law of thermodynamics, assumes the form

$$
dQ = \frac{1}{\mu} dS_{3/\alpha} = T \frac{S_{3/\alpha}^{\alpha/3-1}}{\kappa} dS_{3/\alpha} = \frac{3T}{\kappa\alpha} dS_{3/\alpha}^{\alpha/3}.
\tag{26}
$$

We will denote $(3S_{3/\alpha}^{\alpha/3}/\kappa\alpha)$ as $\mathcal{S}_{(\alpha)}$ and note that $\mathcal{S}_{(\alpha)} \propto L^\alpha$. By analogy with (6), the proportionality factor between $\mathcal{S}_{(\alpha)}$ and $L^\alpha$ in the limit of a large $L$ is set to be $(4\pi)^{\alpha/2}\gamma_{\alpha/2}$, where the value of $\gamma_{\alpha/2}$ is still to be determined (see the next sub-section). Finally, we can express the first law of thermodynamics in a simple form as

$$dU = Td\mathcal{S}_{(\alpha)} - pdV. \tag{27}$$

This can also be obtained with the help of the zeroth law of thermodynamics (cf. Appendix A). We note that, similar to fluid dynamics, the work density $W$ plays the role of pressure in the cosmological framework. So $pdV \rightarrow WdV$. In particular, in cosmology, $W = -\frac{1}{2}\text{Tr}(T^{\mu\nu})$, where "Tr" denotes the two-dimensional normal trace, i.e., $\text{Tr}(T^{\mu\nu}) = T^{\alpha\beta}h_{\alpha\beta}$. In the latter, $T^{\mu\nu}$ is the energy–momentum tensor, and $h_{\alpha\beta}$ represents the metric on the horizon [24,41].

### 2.2. $\mathcal{S}_\delta$-entropy and Friedmann Equations

Equation (27) is often employed as a basis for deriving cosmological equations in Tsallis cosmology, even though the justification and rationale for its use are based on a premise that is distinct from that discussed in the preceding sub-section. In the literature, Equation (27) is typically derived from a formal analogy with black hole thermodynamics [70,**?** ]. In contrast, our subsequent approach directly follows the original Tsallis proposal, which considers cosmological systems with holographic micro-state-space scaling as genuine thermodynamic systems, provided that an extensive, non-additive entropy $S_\delta$ is used.

By utilizing Equation (27), one can deduce the first and second modified Friedmann equations for a homogeneous and isotropic universe. In particular, for the flat FRW universe, these read (see, e.g., Ref. [41])

$$\frac{8\pi M_{\text{Pl}}^{-\alpha}}{3}\rho = \left(H^2\right)^{2-\alpha/2}, \tag{28}$$

$$\frac{\ddot{a}}{a}\left(H^2\right)^{1-\alpha/2} = \frac{8\pi M_{\text{Pl}}^{-\alpha}}{3(4-\alpha)}[(1-\alpha)\rho - 3p]. \tag{29}$$

Here, $a(t) > 0$ is the Robertson–Walker scale factor, $H = \dot{a}/a$ is the Hubble parameter, and $M_{\text{Pl}}$ is the Planck mass [in natural units ($c = 1 = \hbar$) $M_{\text{Pl}} = \sqrt{1/G} \simeq 2.18 \times 10^{19}$ GeV $\simeq 2.18 \times 10^{-8}$ kg]. The function $\rho$ represents the energy density of the universe's matter content. Since the FRW universe must have the perfect fluid form, $\rho$ is the perfect fluid energy density. In addition, the proportionality parameter $\gamma_{\alpha/2}$ assumes the form [41]

$$\gamma_{\alpha/2} = \frac{3(4-\alpha)(4\pi)^{1-\alpha/2}M_{\text{Pl}}^\alpha}{4\alpha}. \tag{30}$$

In connection with (29), one interesting observation is in order. Cosmological measurements, including type Ia supernovae [71], the cosmic microwave background (CMB) [72], and the large-scale structure [73,74], suggest that the universe is currently in an accelerated phase, which means that $\ddot{a} > 0$. By using the equation of state for a perfect fluid $p = w\rho$ (where $w$ is a dimensionless number), we obtain from (29) that

$$(1-\alpha)\rho - 3w\rho > 0 \quad \Rightarrow \quad w < (1-\alpha)/3. \tag{31}$$

We have used the fact that $\alpha$ is less than 4 to obtain Inequality (31). In fact, $\alpha$ can be maximally 3 (as seen in the Introduction). We note that (31) implies that for $\alpha \geq 1$, $w$ is always negative (thus corresponding to dark energy), while for $\alpha < 1$, $w$ can be either negative or positive. This means that in Tsallis cosmology, the accelerated phase of the late-time universe is possible even with ordinary matter, i.e., without invoking the concept of dark energy. In particular, for $w = 0$ (ordinary dust matter), accelerated expansion can be obtained with the scaling exponent $\alpha < 1$. In this paper, we will, however, not be exploring any further the fascinating topic of the universe's accelerated phase. This is because our focus here will be on a radiation-dominated universe, which is the early-universe era whose dynamics is determined by radiation, such as photons, neutrinos and ultra-relativistic electrons and positrons.

To proceed, we rewrite (28) as

$$H(T) = Q(T) H_{\text{St.Cosm.}}(T),$$  (32)

where $H_{\text{St.Cosm.}} = \sqrt{\dfrac{8\pi}{3M_{\text{Pl}}^2} \rho(T)}$ is the Hubble parameter in the standard cosmology and $Q(T)$ (the so-called amplification factor) is given by

$$Q(T) = \left[ \sqrt{\frac{8\pi}{3}} \frac{\rho^{1/2}}{M_{\text{Pl}}^2} \right]^{\frac{\alpha-2}{4-\alpha}} = \eta \left( \frac{T}{T_*} \right)^{\nu},$$  (33)

where

$$\eta = \left[ \frac{2\pi}{3} \sqrt{\frac{\pi g_*(T)}{5}} \right]^{\frac{\alpha-2}{4-\alpha}},$$  (34)

$$\nu = \frac{2(\alpha-2)}{4-\alpha}, \quad T_* \equiv M_{\text{Pl}}.$$  (35)

In deriving (33) we used the fact that for a radiation-dominated universe, the Stefan–Boltzmann law $\rho = \frac{\pi^2 g_*(T)}{30} T^4$ holds. In the latter, $g_*(T)$ counts the total number of effective degrees of freedom (those species with the rest mass $m_i \ll T$), cf. [75]. The explicit form of the amplification factor (33) is the key input from Tsallis cosmology, and it will be crucial in the following two sections.

Another important consequence of the generalized Friedman equations is the modified time scaling for $a(t)$. To see this, we observe that the equation of state for radiation, $p = \rho/3$, implies the continuity equation

$$\dot{\rho}(t) + 4H\rho(t) = 0.$$  (36)

This is a consequence of the conservation of the energy–momentum tensor in the FRW background, i.e., $\nabla_\mu T^{\mu\nu} = 0$. Equation (36) can be solved with $\rho(t) = \rho_0/a^4(t)$, where $\rho_0$ is a constant. By inserting this relation into Equation (28), we obtain that the scale factor $a(t) = a_0 \left( \frac{t}{4-\alpha} \right)^{1-\alpha/4}$, which should be compared to the scaling behavior $a(t) \propto t^{1/2}$ that results from the standard Friedmann equations. In addition, by employing the Stefan–Boltzmann law, we obtain the relation between the cosmic time $t$ and the temperature $T$, namely $t \propto T^{\frac{4}{\alpha-4}}$, which implies that $Ta(t) = \text{constant}$.

## 3. Tsallis Cosmology and Bounds from BBN

### 3.1. General Analysis

In this section, we will examine the effects of Tsallis cosmology, discussed in the previous section, on Big Bang nucleosynthesis. In Appendix B, we provide supplementary technical details regarding the derivation of the equations employed here. In our exposition, we will chiefly follow the approach discussed in Bernstein et al. [76], Torres et al. [77], and Capozzielo et al. [78].

We start by equating the expansion rate of the universe (32) with the interaction rates of relevant processes involved during the BBN (cf. Appendix B). This allows us to compute the freeze-out temperature $T_f$ [1]

---

[1] Before the nucleosynthesis epoch, neutrons and protons interconverted between themselves and maintained thermal equilibrium via weak interaction reactions, such as $n + \nu_e \leftrightarrow p + e^-$. Thermalizing reactions such as these stopped being effective at maintaining a thermal particle distribution when the decreasing temperature

$$T_f = M_{\text{Pl}} \left[ \eta \frac{2\pi}{3} \sqrt{\frac{\pi g_*}{5}} \frac{1}{\mathcal{A}_0 M_{\text{Pl}}^4} \right]^{\frac{1}{3-\nu}}, \tag{37}$$

where $\mathcal{A}_0 = 9.6 \times 10^{-10}$ GeV$^{-4}$. Defining $\delta T_f = T_f - T_{0f}$, with $T_{0f} \sim 0.76$ MeV (which follows from the standard computation with $H_{\text{St.Cosm.}} \simeq \mathcal{A}_0 T^5$), one obtains

$$\left| \frac{\delta T_f}{T_f} \right| = \left| 1 - \frac{T_{0f}}{M_{\text{Pl}}} \left[ \left( \frac{2\pi}{3} \sqrt{\frac{\pi g_*}{5}} \right)^{\frac{2}{4-\alpha}} \frac{1}{\mathcal{A}_0 M_{\text{Pl}}^4} \right]^{-\frac{4-\alpha}{16-5\alpha}} \right|. \tag{38}$$

At the same time, the BBN consistent $^4He$ mass fraction [79,80]

$$Y_p = 0.2449 \pm 0.004, \tag{39}$$

with an uncertainty $|\delta Y_p| < 10^{-3}$ can be used to infer an upper bound on (38). In particular, from Equation (A18), one can directly deduce that

$$\left| \frac{\delta T_f}{T_f} \right| < 3.7 \times 10^{-2}. \tag{40}$$

The constraint on the parameter $\alpha$ can be obtained by comparing Equations (38) and (40). For $T_f \simeq 0.76$ MeV, we have to use $g_* \simeq 10$. In fact, in the standard cosmology, the value $g_* \simeq 10$ is constant in the temperature range 1 MeV $\lesssim T \lesssim 100$ MeV, cf. [75]. Figure 1 shows that BBN restricts the values of $\alpha$ to the interval $2.0013 \lesssim \alpha \lesssim 2.0057$. The later means that

$$1.4957 \lesssim \delta \lesssim 1.4990 \quad \Leftrightarrow \quad 0.0013 \lesssim \Delta \lesssim 0.0057. \tag{41}$$

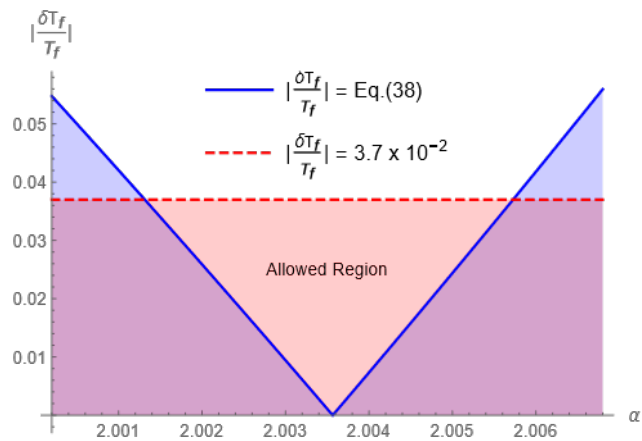

**Figure 1.** $|\delta T_f/T_f|$ vs $\alpha$. Here, $\delta T_f/T_f$ is defined in (38), while the upper bound for $\delta T_f/T_f$ is given in Equation (40). The BBN provides values for $\alpha$ in the range $2.0013 \lesssim \alpha \lesssim 2.0057$. This means that Tsallis' $\delta$ is in the range $1.4957 \lesssim \delta \lesssim 1.4990$.

### 3.2. Constraints on Tsallis Cosmology from Primordial Abundance of Light Elements

Here, the bounds on the Tsallis parameter $\delta$ are derived by analyzing the primordial abundances of light elements, namely deuterium $^2H$, helium $^4He$, and lithium $^7Li$. Since the uncertainties in respective mass abundances are different from that used in the previous

---

and density induced a slowing down of the reactions with respect to the expansion rate of the universe. Such a deviation from equilibrium led to the *freeze-out* of massive particles. The onset temperature at which this happened is known as the *freeze-out temperature*.

section, the bounds on $\delta$ will also be slightly different. The $Q$-term entering the primordial light elements [81] is replaced by the amplification factor (33) (where the value $Q \neq 1$ corresponds to the modification of GR induced by Tsallis cosmology). Moreover, we shall assume three generations of neutrinos ($N_\nu = 3$). For the following analysis, it is important to recall the main features of the formation of light elements. Following Refs. [82,83], we have:

- $^4He$ *abundance*—Helium $^4He$ production is generated by deuterium $^2H$ production by means of a neutron and a proton. This is then converted into $^3He$ and tritium (the relevant reactions we are considering are $n + p \rightarrow {}^2H + \gamma$, $^2H + {}^2H \rightarrow {}^3He + n$, and $^2H + {}^2H \rightarrow {}^3H + p$. Helium $^4He$ is produced by the reactions $^2H + {}^3H \rightarrow {}^4He + n$; $\;{}^2H + {}^3He \rightarrow {}^4He + p$.). The best fit for the primordial $^4He$ abundance is given by [84,85]

$$Y_p = 0.2485 \pm 0.0006 + 0.0016[(\eta_{10} - 6) + 100(Q - 1)], \qquad (42)$$

where $Q$ is the amplification factor (in our case described by (33)) and $\eta_{10}$ (the *baryon density parameter*) is defined as $\eta_{10} \equiv 10^{10}\eta_B$; see, e.g., [86,87], where $\eta_B = n_B/n_\gamma$ is the *baryon-to-photon ratio* [88]. The values $\eta_{10} = 6$ and $Q = 1$ correspond to the standard BBN result for the $^4He$ mass fraction based on the standard cosmological model (St.Cosm.), which yields $(Y_p)|_{\text{St.Cosm.}} = 0.2485 \pm 0.0006$. Using the observational data [89] and $\eta_{10} = 6$ (see, e.g., [86,87]), one obtains the correct phenomenological helium $^4He$ mass abundance $0.2449 \pm 0.0040$, cf. Equation (39), provided

$$0.2449 \pm 0.0040 = 0.2485 \pm 0.0006 + 0.0016[100(Q - 1)]. \qquad (43)$$

This relation provides the sought constrain on $Q$. By taking $Q$ to be equal to $Q_{4He}$ (see, e.g., Ref. [82]), we obtain $Q_{4He} = 1.0475 \pm 0.105$. Figure 2 displays the latter relation with $Q$ arising from the Tsallis cosmology, cf. Equation (33). This also imposes a constraint on admissible values of $\alpha$. The permissible range of $\alpha$'s in relation to helium abundance is

$$1.9967 \lesssim \alpha \lesssim 2.0014$$

$$\Rightarrow \quad 1.4989 \lesssim \delta \lesssim 1.5024 \quad \Leftrightarrow \quad -0.0033 \lesssim \Delta \lesssim 0.0014 \,. \qquad (44)$$

- $^2H$ *abundance*—Neutron–proton interactions, that is, $n + p \rightarrow {}^2H + \gamma$, produce deuterium, $^2H$. Presently, the best fit for deuterium abundance is given by [86]

$$Y_{2H} = 2.6(1 \pm 0.06)\left(\frac{6}{\eta_{10} - 6(Q - 1)}\right)^{1.6}. \qquad (45)$$

The values of $Q = 1$ and $\eta_{10} = 6$ once again result in a standard cosmology with the value of $Y_{Dp}|_{\text{St.Cosm.}} = 2.6 \pm 0.16$. The observational constraint on deuterium abundance $Y_{Dp} = 2.55 \pm 0.03$, cf. Ref. [89] and Equation (45), imply

$$2.55 \pm 0.03 = 2.6(1 \pm 0.06)\left(\frac{6}{\eta_{10} - 6(Q - 1)}\right)^{1.6}, \qquad (46)$$

leading to the constraint on $Q \equiv Q_{2H}$ given by $Q_{2H} = 1.062 \pm 0.444$. Following the same strategy as in the helium abundance case, we compare the latter with the amplification factor (33). The result is reported in Figure 2. From this, we can deduce the corresponding range of variability of $\alpha$ for $^2H$ abundance, namely

$$1.9906 \lesssim \alpha \lesssim 2.011$$

$$\Rightarrow \quad 1.4918 \lesssim \delta \lesssim 1.5071 \quad \Leftrightarrow \quad -0.0094 \lesssim \Delta \lesssim 0.011 \,. \qquad (47)$$

- $^{7}Li$ *abundance*—When considering lithium abundance, the $\eta_{10}$ parameter successfully fits the abundances of $^{2}H$ and $^{4}He$, but it does not align with the observations of $^{7}Li$. This fact is known as the *lithium problem* [81]. In standard cosmology, the ratio of the expected value of $^{7}Li$ abundance and the observed value (Obs.) is $(Li|_{\text{St.Cosm.}})/(Li|_{\text{Obs.}}) \in [2.4, 4.3]$, cf., e.g., [81,90]. The best fit for $^{7}Li$ abundance is presently given by [86], namely

$$Y_{Li} = 4.82(1 \pm 0.1)\left[\frac{\eta_{10} - 3(Q - 1)}{6}\right]^{2}. \tag{48}$$

The phenomenological constraint on lithium abundance $Y_{Li} = 1.6 \pm 0.3$, cf. Ref. [89], yields $Q_{Li} = 1.960025 \pm 0.076675$, see [82].

Such a value does not overlap with the constraints on $^{2}H$ and $^{4}He$ abundances. In fact, from Figure 2, we see that the range for admissible $\alpha$s is

$$1.9836 \lesssim \alpha \lesssim 1.9854$$

$$\Rightarrow \quad 1.511 \lesssim \delta \lesssim 1.5124 \quad \Leftrightarrow \quad -0.0164 \lesssim \Delta \lesssim -0.0146. \tag{49}$$

The above results indicate that there is no overlap between $\alpha$s from $\{^{4}He, ^{2}H\}$ and $^{7}Li$, i.e., $\alpha$s from Equations (44), (47), and (49), respectively. Therefore, the lithium problem persists also in Tsallis cosmology. On the other hand, the respective ranges of admissible $\alpha$s (and $\delta$s) are tantalizingly close. This indicates that within the context of Tsallis cosmology, the lithium problem could potentially be mitigated by considering additional elements of $\delta$-entropy statistics, for example, considering the effects of the $\delta$-entropy-based statistics (the $\delta$-entropy-based statistics is a statistical physics based on a maximum entropy probability distribution) on the computation of the relevant (electroweak) reactions occurring during the BBN. Such a study is, however, beyond the scope of the present paper.

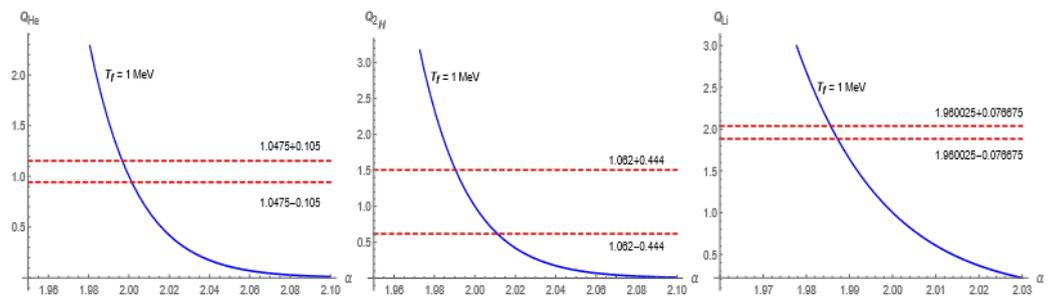

**Figure 2.** $\{Q_{^{4}He}, Q_{^{2}H}, Q_{^{7}Li}\}$ vs $\alpha$. The experimental ranges of $Q_{^{4}He, ^{2}H, ^{7}Li}$ are reported. The baryon parameter is fixed to $\eta_{10} = 6$, while the freeze-out temperature is set as $T_{f} \simeq 1$ MeV.

## 4. Tsallis Cosmology and Bounds from the Relic Abundance of Cold Dark Matter Particles

Further constraint on $\alpha$ can be deduced by utilizing the dark matter (DM) annihilation cross-section, which is linked to the cold DM relic abundance $\Omega_{\text{CDM}}$. In order to achieve this, we connect Tsallis cosmology with the $f(R) = M_{*}R^{n}$ cosmology following the approach outlined in Ref. [91]. It should be noted that the scenario where $n = 1$ corresponds to the conventional Einsteinian cosmology. The aforementioned strategy will enable us to utilize the results derived in [92].

In a spatially flat FRW metric, and under the assumption that the scale factor follows a power-law evolution, $a(t) = a_{0}t^{\varsigma}$ (in the case of Tsallis cosmology $\varsigma = 1 - \alpha/4$, as discussed in Section 2.2), the $f(R)$ cosmological equations can be expressed in the form that is formally identical to (32) provided the amplification factor $Q(T)$ is substituted with $Q_n(T)$, namely (cf. Refs. [93,94])

$$Q(T) \rightarrow Q_n(T) = \eta_n \left( \frac{T}{M_{Pl}} \right)^{\nu_n},$$ (50)

where

$$\eta_n = \sqrt{\frac{\gamma(8\pi/3)^{(1-n)/n}}{6|2\varsigma - 1|}} \left( \frac{\pi^2 g_*}{30} \right)^{\frac{1-n}{2n}} \frac{1}{(\tilde{M}_* \Omega)^{1/2n}},$$ (51)

$$\nu_n = \frac{2}{n} - 2.$$ (52)

Here, the dimensionless constant $\tilde{M}_*$ is linked to the constant $M_*$ by the equation $\tilde{M}_* = M_* M_{Pl}^{-2(1-n)}$, while $\Omega$ is related to $n$ [93], but its explicit expression is not relevant for us. Comparing (35) and (52), one obtains

$$n = 2 - \frac{\alpha}{2} = 1 - \frac{\Delta}{2} = 2 - \frac{3}{2\delta},$$ (53)

while by comparing (33) with (51), we obtain the relation between $M_*$ and the parameters $\{\alpha, \varsigma\}$. Being irrelevant to our analysis, we will not report the relation here.

These results can be used for investigating DM relic abundance (we assume that the DM is composed of weakly interacting massive particles (WIMPs) that are conventionally assumed to be fermions). In modified cosmology, the cold DM relic density assumes the form [75,92]

$$\Omega_{\text{CDM}} h^2 \simeq 10^9 \frac{(\bar{l} + 1) x_f^{(\bar{l}+1)}}{(h_*/g_*^{1/2}) M_{Pl} \bar{\sigma}},$$ (54)

where $h_*$ is the number of relativistic degrees of freedom for entropy density (typically $h_* \sim g_*$) and $x_f \equiv m/T_f$ has an explicit form

$$x_f = \log[0.038(\bar{l} + 1)(g/g_*^{1/2}) M_p m \bar{\sigma}]$$
$$- (\bar{l} + 1) \log\{\log[0.038(\bar{l} + 1)(g/g_*^{1/2}) M_p m \bar{\sigma}]\}.$$ (55)

Here, $g = 2$ is the spin polarization of the DM particle, $m$ is the mass of the WIMP particle, $\bar{\sigma}$ is the WIMP cross-section, $\bar{l} = l + (1 - n)$, and $h_*$ is the number of relativistic degrees of freedom for entropy density [92]. Here, $\bar{l} = l$ for conventional GR [2] (where, as we know, $\alpha = 2$), while $l = 0, 1$ correspond, respectively, to $s$-wave and $p$-wave polarizations. Cosmological observations constrain the cold DM density to $\Omega_{\text{CDM}} h^2 = 0.1198 \pm 0.0012$ (see Ref. [72]), where $h \in [0.2, 1]$ is the reduced Hubble constant [75]. Following the analysis of Ref. [92], one finds that the annihilation of the cross-section ($\sigma/(10^{10} \text{GeV}^{-1}/M_{Pl})$) vs. allowed WIMP masses ($m(\text{GeV}) \in [10^2, 5 \times 10^2]$) for the cold DM abundance $\Omega_{\text{CDM}}$ gives that $n \in (1-10^{-4}, 1-10^{-3})$. In obtaining the latter bound, we have used in (54) the relation (cf., e.g., [75])

$$\bar{\sigma} = \frac{3.2 g_*^{1/2+/2n}}{n} \left( \frac{m}{M_P} \right)^{2-2/n} \left( \frac{4\pi^3 g_*}{15} \right)^{-2/n} \sigma.$$ (56)

---

[2] Notice that in [92], it has been used the parametrization $\langle \sigma v \rangle = \sigma_0 x^{-l}$, where $l = 0$ corresponds to $s$-wave annihilation, $l = 1$ to $p$-wave annihilation, and so on. The modification of standard cosmology induces the corrections to the parameter $l$ via $\bar{l} = l + (1 - n)$. In the case of GR, $n = 1$, and one obtains $\bar{l} = l$.

From Equation (53), we thus see that $\alpha$ has the range of admissibility

$$2.0002 \lesssim \alpha \lesssim 2.01 \quad \Rightarrow \quad 1.493 \lesssim \delta \lesssim 1.499 \quad \Leftrightarrow \quad 0.0002 \lesssim \Delta \lesssim 0.01 \,. \tag{57}$$

The key takeaway from this section is that even a minor deviation of Tsallis cosmology from the standard cosmological model can have significant impacts on the cross-section of WIMPs. Remarkably, present values of $\alpha$, $\delta$, and $\Delta$ are also compatible with the values (41), (44), and (47) obtained in the framework of the BBN.

## 5. Discussion and Conclusions

This paper delves into the thermodynamic structure implied by Tsallis' $\delta$-entropy, emphasizing the crucial role played by an integrating factor of the heat one-form. In particular, our emphasis was on the precise formulation of the first law of thermodynamics. Furthermore, the zeroth law of thermodynamics has also been addressed, and the role of empirical temperature in determining absolute temperature has been elucidated.

With the first law of thermodynamics at hand, we have addressed the issue of Tsallis cosmology and its prospective role in the Big Bang nucleosynthesis and the relic abundance of cold DM particles. From the perspective of BBN, we have determined the bounds on the Tsallis parameter $\alpha$ through both a general analysis of BBN using the freeze-out temperature formula and an examination of the current best fits for primordial abundances of light elements $^4He$, $^2H$, $^7Li$. The admissible range of $\alpha$s steaming from freeze-out temperature variation is given in Equation (41), while from the primordial abundance of light elements, we deduced the bounds (44), (47), and (49). As noted, there is a pairwise overlap of the ranges of $\alpha$ originating both from the freeze-out temperature formula and from helium and deuterium abundances with values around $\alpha \simeq 2.0013$, while for lithium, such an overlap does not exist (the lithium problem). On the other hand, the range of admissible $\alpha$s from lithium is tantalizingly close to admissible values from other considered BBN sources. This indicates that within the context of Tsallis cosmology, the lithium problem could potentially be mitigated by considering additional elements of $\delta$-entropy statistics. At this point, it should be noted that all that was needed to pass from the first law of thermodynamics to this modified cosmology was the entropy scaling law (9). In particular, the derivation of modified Friedman equations did not require explicit knowledge of the entropic functional (i.e., how it depends on probability). Thus, by going beyond such a paradigm, one could, for instance, utilize the effects of the $\delta$-entropy-based statistics on the computation of the relevant (electroweak) reactions occurring during the BBN. Such a statistical physics consideration would, however, go beyond the scope of the present paper. As a next point, we have investigated the role of Tsallis cosmology in the framework of cold DM theory. We have found that the tiny deviation from the standard cosmological scenario induced by $\alpha$-corrections may account for the observed relic dark matter abundance, with a value of the Tsallis parameter $\alpha$ compatible with the constraints obtained from BBN.

In passing, we might note that the obtained values $\alpha \simeq 2$ are not compatible with the bound $\alpha < 1$, i.e., a situation which allows us to explain the accelerated phase of the late-time universe without invoking the concept of dark energy. On the other hand, one might assume, in the spirit of renormalization theory, that the anomalous dimension $\Delta$ runs during the evolution of the universe from the BBN era to today, so that at low energies $\Delta < -1$ ("porous" horizon surfaces), or, in other words, $\alpha_{\mathrm{BBN}} \simeq 2 \to \alpha_0 < 1$, where the index 0 refers to the current value of $\alpha$. A scenario along these lines has been recently considered, e.g., in Ref. [58].

The results discussed in this paper can be further employed in various ways. In particular: (1) they might contribute to the ongoing debate on the most pertinent cosmological scenario among models based on Tsallis' $\delta$-entropy (and hence the ensuing entropic origin of gravity) and the dark sector; (2) they might be instrumental in inferring bounds on Tsallis cosmology from primordial gravitational waves. The latter may be successfully addressed provided the tensor perturbations, generated during the inflation era and propagated during the Tsallis cosmological era, will be clear enough to be measurable by the future gravitational-wave detectors [95]. Work along those lines is presently being actively pursued.

**Author Contributions:** Conceptualization, P.J. and G.L.; formal analysis, G.L.; methodology, P.J. and G.L.; software design, data structures, computer calculation, and visualization, G.L.; writing—original draft, P.J.; writing—review and editing, P.J. and G.L. All authors have read and agreed to the published version of the manuscript.

**Funding:** P.J. was in part supported by the Ministry of education grant MŠMT RVO 14000.

**Institutional Review Board Statement:** Not applicable.

**Informed Consent Statement:** Not applicable.

**Data Availability Statement:** Not applicable.

**Conflicts of Interest:** The authors declare no conflict of interest.

## Abbreviations

The following abbreviations are used in this manuscript:

| | |
|---|---|
| BH | Bekenstein–Hawking |
| GR | General relativity |
| QFT | Quantum Field Theory |
| BBN | Big Bang nucleosynthesis |
| FRW | Friedmann–Robertson–Walker |
| CMB | Cosmic Microwave Background |
| DM | Dark Matter |
| WIMP | Weakly-interacting massive particle |
| CDM | Cold dark matter |

## Appendix A. Zeroth Law of Thermodynamics and $S_\delta$ Entropy

In this appendix, we discuss the zeroth law of thermodynamics for systems described by $S_\delta$ entropy, which complements Section 2.1. The zeroth law of thermodynamics codifies the concept of *thermal equilibrium* and posits that if two thermodynamic systems are in thermal equilibrium with each other, and also separately in thermal equilibrium with a third system, then the three systems are in thermal equilibrium with each other. This transitive property of thermal equilibrium allows to divide the Universe into disjoint classes of systems that are in thermal equilibrium with each other, and to quantify each such class with a unique number known as *empirical temperature*, i.e., physical temperature (such as the Celsius scale) which may not not necessarily coincide with the absolute temperature. Aside from thermal equilibrium systems can also be in another type of equilibria, e.g., in mechanical, chemical or diffusive equilibria. For example, *physical pressure* quantifies the mechanical equilibrium of homogeneous chemical systems in physical contact.

To address the issue of the zeroth law of thermodynamics, we start by considering two systems ($A$ and $B$) in contact (both thermal and mechanical) with each other. Suppose that these have volumes $V(A)$ and $V(B)$ and internal energies $U(A)$ and $U(B)$, and that the total internal energy and total volume are fixed.

In thermodynamic equilibrium, the total entropy $S_\delta(A+B)$ must be maximal. By using (14) we thus have

$$
\begin{aligned}
0 \;=\; & dS_\delta(A+B) \;=\; d\Big[S_\delta^{1/\delta}(A) \;+\; S_\delta^{1/\delta}(B)\Big]^\delta \\[2mm]
=\; & [S_\delta(A+B)]^{1-1/\delta}\Bigg\{[S_\delta(A)]^{1/\delta-1}\left(\frac{\partial S_\delta(A)}{\partial U(A)}\right)_{V(A)} \\[2mm]
& \qquad\qquad -\,[S_\delta(B)]^{1/\delta-1}\left(\frac{\partial S_\delta(B)}{\partial U(B)}\right)_{V(B)}\Bigg\}dU(A) \\[2mm]
& +\,[S_\delta(A+B)]^{1-1/\delta}\Bigg\{[S_\delta(A)]^{1/\delta-1}\left(\frac{\partial S_\delta(A)}{\partial V(A)}\right)_{U(A)} \\[2mm]
& \qquad\qquad -\,[S_\delta(B)]^{1/\delta-1}\left(\frac{\partial S_\delta(B)}{\partial V(B)}\right)_{U(B)}\Bigg\}dV(A)\,, \quad (A1)
\end{aligned}
$$

where we have employed the fact that the total internal energy and volume are fixed

$$
U(A+B) \;=\; U(A) \;+\; U(B) \;=\; \text{const.}\,, \tag{A2}
$$

$$
V(A+B) \;=\; V(A) \;+\; V(B) \;=\; \text{const.}\,. \tag{A3}
$$

We have also assumed that the $S_\delta$ entropy is expressed in terms of its natural state variables, namely $U$ and $V$.

From (A1), we obtain two identities that reflect the simultaneous thermal and mechanical equilibrium of systems $A$ and $B$. The first identity can be expressed as follows:

$$
k_\delta\beta(A)\,[S_\delta(A)]^{1/\delta-1} \;=\; k_\delta\beta(B)\,[S_\delta(B)]^{1/\delta-1} \;\equiv\; k_\delta\beta^*\,, \tag{A4}
$$

where (by analogy with conventional thermodynamics) we have defined

$$
k_\delta\beta \;=\; \left(\frac{\partial S_\delta}{\partial U}\right)_V\,. \tag{A5}
$$

It is important to emphasise that the physical temperature is not equal to $(k_\delta\beta)^{-1}$, but rather:

$$
\vartheta \;=\; \frac{1}{k_\delta\beta^*} \;=\; \frac{[S_\delta(B)]^{1-1/\delta}}{k_\delta\beta}\,. \tag{A6}
$$

Equation (A4) encapsulates the zeroth law of thermodynamics, which guarantees that the same empirical temperature $\vartheta$ can be assigned to all subsystems in thermal equilibrium.

The second identity can be cast as

$$
\begin{aligned}
\left(\partial S_\delta^T(A)/\partial V(A)\right)_{U(A)}&[S_\delta(A)]^{1/\delta-1} \\[2mm]
&= \left(\partial S_\delta^T(B)/\partial V(A)\right)_{U(B)}[S_\delta(B)]^{1/\delta-1} \;\equiv\; \frac{p_{\text{phys}}}{\vartheta}\,. \quad (A7)
\end{aligned}
$$

Equation (A7) reflects that when two systems are in mechanical equilibrium, their pressures are equal. This allows to identify *physical pressure* $p_{\text{phys}}$ as

$$
p_{\text{phys}} \;=\; \vartheta[S_\delta(B)]^{1/\delta-1}\left(\frac{\partial S_\delta}{\partial V}\right)_U\,. \tag{A8}
$$

When the microstates scale exponentially with volume (as in conventional thermodynamics) then $\delta = 1$, cf. Equations (1) and (2). In fact, in the limit $\delta \rightarrow 1$ Equation (A8) approaches the conventional result.

Note that by writing [as in Equations (A6) and (A8)]

$$
\begin{aligned}
dS_\delta &= \left( \frac{\partial S_\delta}{\partial U} \right)_V dU + \left( \frac{\partial S_\delta}{\partial V} \right)_U dV \\
&= [S_\delta]^{1-1/\delta} \frac{1}{\vartheta} dU + [S_\delta]^{1-1/\delta} \frac{p_{\text{phys}}}{\vartheta} dV ,
\end{aligned}
\tag{A9}
$$

we obtain

$$
\frac{\vartheta}{[S_\delta]^{1-1/\delta}} dS_\delta = dU + p_{\text{phys}} dV .
\tag{A10}
$$

Since

$$
dS_\delta = \mu \, d\mathcal{Q} = \frac{[S_\delta]^{1-1/\delta} \kappa}{T} \, d\mathcal{Q} \quad \Rightarrow \quad \frac{\vartheta}{[S_\delta]^{1-1/\delta}} dS_\delta = \frac{\vartheta \kappa}{T} \, d\mathcal{Q} .
\tag{A11}
$$

We note that the first law of thermodynamics follows from (A10) if we equate the empirical temperature defined in (A6) with the absolute temperature $T$ discussed in Section 2.1. For $\vartheta$ so chosen, we obtain that $w(\vartheta)$ from (24) must be $1/\vartheta$. Additionally, $T$ has the same units as the temperature $\vartheta$, provided $\kappa = 1$. Consequently, the ensuing first law of thermodynamics reads

$$
d\mathcal{Q} = T \frac{[S_\delta]^{1/\delta - 1}}{\kappa} dS_\delta = dU + p_{\text{phys}} dV ,
\tag{A12}
$$

which precisely coincides with the form (27). When we require the Legendre structure, then we must set $\delta = 3/\alpha$ for the microstate scaling (9).

## Appendix B. BBN Physics—A Short Review

In this appendix, we provide an overview of some essentials of Big Bang nucleosynthesis that are needed in Section 3. Let us first recall that BBN occurred in the early Universe when the temperature was $T \sim \mathcal{O}(1)$ MeV. During this epoch, the energy density was dominated by electrons, positrons, neutrinos and photons, which were in thermal equilibrium owing to the weak interaction neutron-proton conversion processes : $e^+ + n \leftrightarrow p + \bar{\nu}_e$, $\nu_e + n \leftrightarrow p + e^-$, and $n \leftrightarrow p + e^- + \bar{\nu}_e$ (the other neutrino flavors do not contribute to these reactions). The conversion rate of protons into neutrons

$$
\Gamma_{np}(T) = \Gamma_{n + \nu_e \rightarrow p + e^-} + \Gamma_{n + e^+ \rightarrow p + \bar{\nu}_e} + \Gamma_{n \rightarrow p + e^- + \bar{\nu}_e} ,
\tag{A13}
$$

and its inverse $\Gamma_{pn}(T) = e^{\mathcal{Q}/T} \Gamma_{np}(T)$ (here $\mathcal{Q} = m_n - m_p = 1.293$MeV is the mass difference between neutron and proton), allow to compute the neutron abundance (see Equation (A15) and details in [75,76]). For instance, the interaction rate for the process $n + \nu_e \rightarrow p + e^-$ is [75,76]

$$
\begin{aligned}
\Gamma_{n + \nu_e \rightarrow p + e^-} &= \int \frac{d^3 p_e}{(2\pi)^3 2E_e} \frac{d^3 p_{\nu_e}}{(2\pi)^3 2E_{\nu_e}} \frac{d^3 p_p}{(2\pi)^3 2E_p} |\mathcal{M}|^2 (2\pi)^4 \\
&\quad \times \delta^{(4)}(p_n + p_{\nu_e} - p_p - p_e) f(E_{\nu_e})[1 - f(E_e)] .
\end{aligned}
\tag{A14}
$$

Here $f(E)$ is the Fermi–Dirac distribution and $\mathcal{M} = \left( \frac{g_w}{8M_W} \right)^2 [\bar{u}_p \gamma^\mu (c_V - c_A \gamma^5) u_n][\bar{u}_e \gamma^\mu (1 - \gamma^5) v_{\nu_e}]$ is the corresponding scattering amplitude ($c_V$ and $c_A$ stand for coupling constants in front of the vector and axial vector interactions, respectively). Similar relationships can be written for the remaining two interaction rates.

The (total) weak interaction rate reads, see, e.g., Refs. [75,76]

$$\Lambda(T) = \Gamma_{np}(T) + \Gamma_{pn}(T) \simeq \mathcal{A}_0 T^5 + \mathcal{O}\left(\frac{\mathcal{Q}}{T}\right), \tag{A15}$$

with $\mathcal{A}_0 = 9.6 \times 10^{-10} \, \text{GeV}^{-4}$. One can estimate the primordial $^4He$ abundance $Y_p$ (referred also mass fraction, that is the quantity of proton number/mass as a fraction of the total proton and neuton numbers/masses) by using the formula [75]

$$Y_p \equiv \lambda \frac{2y(t_f)}{1 + y(t_f)}. \tag{A16}$$

Here $\lambda$ represents the fraction of neutrons decaying into protons in the interval between the freeze-out time of the weak interactions $t_f$ and the freeze-out time of the nucleosynthesis $t_n$. Explicitly, $\lambda$ is given by the relation $\lambda = e^{-(t_n - t_f)/\tau}$ where $\tau = 8803 \pm 1.1$ seconds is the neutron mean lifetime [96]. The function $y(t_f)$ denotes the neutron-to-proton equilibrium ratio at $t_f$, namely

$$y(t_f) = \left(\frac{n}{p}\right)_{\text{freeze-out}} = e^{-\mathcal{Q}/T_f}. \tag{A17}$$

The variation of the freeze-out temperature $T_f$ in (A16) implies a formula for the deviations from the $^4He$ mass fraction

$$\delta Y_p = Y_p \left[ \left(1 - \frac{Y_p}{2\lambda}\right) \log\left(\frac{2\lambda}{Y_p} - 1\right) - \frac{2t_f}{\tau} \right] \frac{\delta T_f}{T_f}. \tag{A18}$$

In (A18), we have set $\delta T(t_n) = 0$ as $T_n$ is fixed by the deuterium binding energy [77,91]. Recent observational data for the primordial $^4He$ mass fraction $Y_p$, see e.g., [79,80], has yielded the numerical value given in (39). By using this together with the fact that $|\delta Y_p| < 10^{-3}$, we can compute an upper bound on $|\delta T_f/T_f|$ from (A18). The estimations for both $\delta$ and $\Delta$ can be then obtained by employing (39) as outlined in Section 3. There we also needed the freeze-out temperature (37), which is computed through the defining relation $\Lambda(T_f) = H(T_f)$ [their expressions are given in (A15) and (32)].